\documentclass[letter,twocolumn]{jpsj2}

\usepackage{bm}

\title{
Indications of Universal Excess Fluctuations in Nonequilibrium Systems
}

\author{
Tatsuro \textsc{Yuge}$^{1,2}$\thanks{E-mail address: yuge@m.tains.tohoku.ac.jp} 
and Akira \textsc{Shimizu}$^{2}$\thanks{E-mail address: shmz@ASone.c.u-tokyo.ac.jp}
}

\inst{
$^{1}$IIAIR, Tohoku University, 
Aoba-ku, Sendai, Miyagi 980-8578
\\
$^{2}$Department of Basic Science, 
University of Tokyo, 
Komaba, Meguro-ku, Tokyo 153-8902
}

\abst{
The fluctuation in 
electric current 
in nonequilibrium steady states is investigated 
by molecular dynamics simulation of macroscopically uniform conductors.
At low frequencies, 
appropriate decomposition of the spectral intensity of current into thermal 
and excess fluctuations provides a simple 
picture of excess fluctuations behaving as shot noise.
This indicates that 
the fluctuation-dissipation relation may be violated 
in a universal manner by the appearance of shot noise 
for a wide range of systems with particle or momentum transport.
}

\kword{nonequilibrium steady state, fluctuation-dissipation relation, 
shot noise, electric conduction}

\begin{document}

\maketitle

In equilibrium states, 
the fluctuation of an observable 
is related {\em universally} to a linear response function 
by the fluctuation-dissipation relation (FDR) \cite{KTH}.
In nonequilibrium steady states (NESSs), 
the FDR is often violated, 
and `excess fluctuation' (XF) appears 
\cite{experiment1,BB00,SU92,BH91,KF,SS1991,Nagaev95,
JB95,Kogan,GMG2005,SUS1993,BMS,book,HK1995,FS1998,UH}.
XF plays crucial roles in many fields of physics, 
including single electron tunneling \cite{UH}, 
the squeezing of photons \cite{HK1995,FS1998}, 
the measurement of fractional charge \cite{KF}, 
and the determination of the fundamental limits 
of quantum interference devices \cite{SS1991}.
However, unlike equilibrium fluctuation, it is not yet well understood 
whether {\em universal} properties exist in XF \cite{FT}.

Experimentally, the FDR violation is hardly observable in heat conduction 
because a convection current or a phase transition 
is induced for large 
temperature difference (which drives heat conduction) 
before XF becomes detectable. 
In contrast, the violation 
has been widely observed
in {\em systems with particle (or momentum) transport}, 
such as electric conductors 
and photoemitting devices
\cite{experiment1,BB00,SU92,BH91,KF,SS1991,Nagaev95,
JB95,Kogan,GMG2005,SUS1993,BMS,book,HK1995,FS1998}.
We therefore consider such systems.

Among such systems are simple systems, 
including mesoscopic conductors 
\cite{experiment1,BB00,SU92,BH91,KF,SS1991,SUS1993,
Nagaev95,JB95,Kogan,GMG2005}, 
conductors with junctions \cite{SUS1993,BMS,book}
(e.g., tunnel and PN junctions), and 
light-emitting diodes \cite{HK1995,FS1998}.
These systems are simple in the sense that 
the number of electron modes is small 
and/or many-body interactions are unimportant 
and/or dissipation is negligible 
and/or the principal origin of XF 
is localized in certain mesoscopic regions.
XF generated in such a case 
takes the form of shot noise
\cite{experiment1,BB00,SU92,BH91,KF,SS1991,Nagaev95,
JB95,Kogan,GMG2005,SUS1993,BMS,book,HK1995,FS1998}.
Here, the term `shot noise' is used in a wide sense, 
which stands for fluctuation 
whose spectral intensity $S_I$ 
is proportional to the absolute value 
of average flux, $\left| \langle I \rangle \right|$ \cite{W_varies}.
The ratio $W$ of $S_I$ to its Poissonian value 
is called the Fano factor, 
which takes various values depending on the details of the systems 
\cite{experiment1,BB00,SU92,BH91,KF,SS1991,Nagaev95,
JB95,Kogan,GMG2005,SUS1993,BMS,book,HK1995,FS1998}.

The situation is completely different for 
{\em uniform macroscopic} conductors, 
for which the assumptions made in 
refs.~\citen{experiment1,BB00,SU92,BH91,KF,SS1991,SUS1993,
Nagaev95,JB95,Kogan,GMG2005,BMS,book,HK1995,FS1998} 
do not hold.
Although the FDR violation is hardly observable in uniform metals, 
it is widely observed in uniform 
semiconductors \cite{book}.
Most experiments on the latter 
showed that XF is dominated by $1/f$ noise, 
which is proportional to $\langle I \rangle^2$ \cite{book}.
Although shot noise 
may also exist in such systems, 
it would be masked by $1/f$ noise \cite{1/f:meso} 
because the latter increases more rapidly 
with increasing $\left| \langle I \rangle \right|$.
However, the origin of $1/f$ noise 
is believed to be imperfections in samples, 
such as the fluctuation in carrier number 
and the migration of impurities, 
which result in a strong sample dependence 
of  $1/f$ noise \cite{book}.
Since imperfections in samples are of secondary interest 
in fundamental physics (nonequilibrium statistical mechanics), 
a natural question arises: 
What fluctuation appears in perfect samples?
In this paper, 
we address this question and report a property of XF 
that may be universal.

The models and results of the previous works on mesoscopic conductors
\cite{BB00,SU92,BH91,KF,SS1991,SUS1993,experiment1,
Nagaev95,JB95,Kogan,GMG2005} 
are not 
applicable to macroscopic conductors, 
because, as mentioned above, many assumptions 
that do not hold in macroscopic conductors 
have been made in those works.
We therefore take a different approach.
That is, 
we use molecular dynamics (MD) simulation 
on a model that we believe 
captures the essential elements 
of macroscopic conductors \cite{YIS,YS_Alder}.
This enables us to study the NESSs of 
perfect samples, without making 
the assumptions made in the works on mesoscopic conductors.
Since we can vary the values of the parameters 
to a great extent, we are able to present results 
that may indicate a universal character.

Except at low temperatures, 
quantum effects seem to play minor roles 
in macroscopic conductors far from equilibrium,
because of the strong decoherence.
We therefore 
use the classical model of electric conduction proposed in ref.~\citen{YIS}, 
which describes doped semiconductors at room temperature well \cite{YS_Alder}.
The system includes three types of classical particles, 
which we call electrons 
(each with mass $m_{\rm e}$ and charge $e$),
phonons (each with mass $m_{\rm p}$), 
and impurities.
Their number densities are denoted by $n_{\rm e}$, $n_{\rm p}$, 
and $n_{\rm i}$, respectively.
For simplicity, we assume a two-dimensional system, 
the size of which is $L_x\times L_y$.
In the $x$-direction, 
we apply an external electric field acting {\em only on electrons},
and impose the periodic boundary condition. 
The boundaries in the $y$-direction
are potential walls for electrons
and thermal walls for phonons.
The thermal walls reflect phonons with 
random velocities sampled from an equilibrium distribution 
with temperature $T_0$ \cite{YIS}.
This enables phonons to carry heat constantly
out of the system 
thereby keeping the system in a NESS.
Impurities are immobile and play a role of providing random potential.

We assume short-range interactions among {\em all} particles.
Since interaction potential is well characterized
by its scattering cross section, 
detailed forms of the potential are expected to be 
irrelevant when studying general 
nonequilibrium properties.
Therefore, we here take a simple form, 
$k_0 (\max \{ 0,d_{jl} \})^{5/2}$.
Here, $k_0$ is a constant and 
$d_{jl}=R_j +R_l -|\mbox{\boldmath$r$}_j-\mbox{\boldmath$r$}_l|$ 
is the overlap of the potential ranges.
$R_j$ is the radius of the potential range
($R_{\rm e}$, $R_{\rm p}$, and $R_{\rm i}$ 
for an electron, phonon, and impurity, respectively), 
and $\mbox{\boldmath$r$}_j$ is  
the position of the $j$-th particle.
We can change the strength of scattering 
by varying $R_j$ (and particle number density). 

This model corresponds to a perfect sample 
because the total number of carriers does not change 
and because impurities do not move.
This system is macroscopically uniform 
although the 
translational invariance is broken 
by impurities and the thermal walls for phonons.
Furthermore, 
the model and results are
also applicable to systems 
that have a mass flow of neutral particles \cite{YS_Alder}. 

We use units in which $m_{\rm e}$, $R_{\rm e}$, $e$, the Boltzmann constant, 
and a reference energy are unity.
Regarding the other parameters, 
the main result, eq.~(\ref{eq:asymptote}), 
is insensitive to their values, as will be shown later.
We here fix $R_{\rm p}=1$, $T_0=1$, and $k_0=4000$; 
the other parameters are varied 
to illustrate the possible universality of the result.

To the investigate nonequilibrium states of this model, 
we perform MD simulation 
using Gear's fifth-order predictor-corrector method \cite{YIS}.
The time-step width is set to $10^{-3}$.
The initial position of each particle is 
randomly arranged so as not to be in contact with the other particles, 
and the initial velocities of the electrons and phonons are given 
by the Maxwell distribution with temperature $T_0$. 
We calculate various quantities 
after the system reaches a NESS.

The electric field applied to the system is 
composed of 
a time-independent field $E$, which is varied in a wide range,
and a time-dependent field $\varepsilon f(t)$, which is small.
The electric field induces electric current 
$I(t) \equiv e n_{\rm e} L_y V^x_{\rm e}(t)$, 
where $V^x_{\rm e}$ is the velocity in the $x$-direction 
(i.e., along the electric field) 
of the center of mass of electrons.
We take $\varepsilon \neq 0$ only when 
we calculate the {\em differential} response function
$\mu (t-\tau;E)$ {\em of a NESS}, which is defined by
\begin{equation}
\langle \delta I(t) \rangle_{E,\varepsilon} 
= \int_{-\infty}^t {\rm d}\tau ~\mu(t-\tau ;E) L_x \varepsilon f(\tau) 
+ O(\varepsilon ^2), 
\end{equation}
for $t >\tau$ and by 
$\mu(t-\tau;E)=0$ for $t<\tau$. 
Here, $\delta I = I - \langle I \rangle_{E,0}$, 
and $\langle \cdots \rangle_{E,\varepsilon}$ denotes 
the average at the NESS 
in the electric field $E+\varepsilon f(t)$.
The convolution theorem yields
$\tilde{\mu}(\omega;E) 
= \lim_{\varepsilon \to 0}
\langle \delta\tilde{I}(\omega) \rangle_{E,\varepsilon} 
/ L_x \varepsilon \tilde{f}(\omega)$, 
where the tilde denotes the Fourier transform.
Note that $\tilde{\mu}(\omega;E)$ differs from that 
in an {\em equilibrium state}, $\tilde{\mu}(\omega;0)$.

\begin{figure}[tb]
\begin{center}
\includegraphics[width=.9\linewidth]{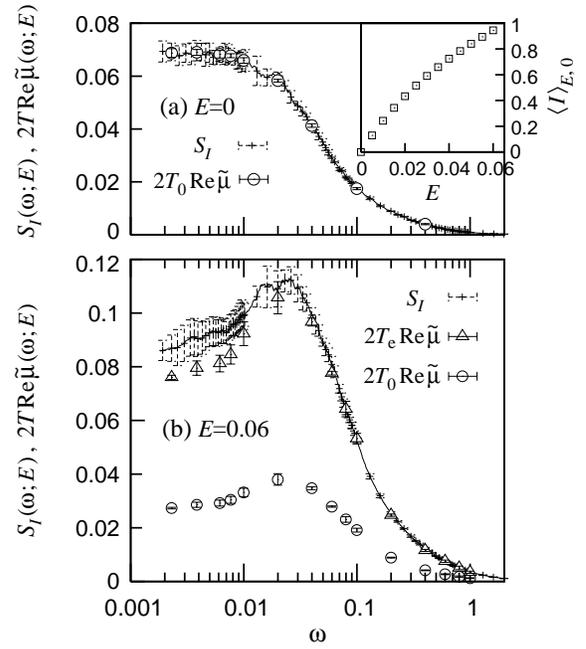}
\end{center}
\caption{
(a) $S_I(\omega;E)$ and $2 T_0 {\rm Re}\tilde{\mu}(\omega;E)$ 
for $E=0$.
The inset shows $\langle I \rangle_{E,0}$ versus $E$.
(b) $S_I(\omega;E)$, $2 T_{\rm e}(E) {\rm Re}\tilde{\mu}(\omega;E)$, 
and $2 T_0 {\rm Re}\tilde{\mu}(\omega;E)$
for $E=0.06$ (nonlinear response regime).
In these simulations, 
$m_{\rm p}=1$, $R_{\rm i}=0.5$, 
$L_x=750$, $L_y=125$, 
$n_{\rm e}=n_{\rm p}=0.016$, and $n_{\rm i}=2/375$.
The data points are the averages of five samples 
(impurity configurations) 
and the error bars are the standard deviations among them.
}
\label{FDR}
\end{figure}

We are mainly interested in the current fluctuation 
that is characterized by the spectral intensity $S_I(\omega;E)$ 
of $I(t)$ for $\varepsilon =0$.
By the Wiener-Khinchine theorem \cite{KTH},
$S_I(\omega;E)$ is equal to the Fourier transform of 
the autocorrelation function 
$\langle \delta I(t) \delta I(0)\rangle_{E,0}$ of current.
In equilibrium states ($E=0$),
the FDR, 
$S_I(\omega;0) = 2 T {\rm Re}\tilde{\mu}(\omega;0)$,
holds for all $\omega$ \cite{KTH}.
Here, $T$ is the temperature of the conductor, 
which is equal to $T_0$ when $E=0$.
We plot both sides of this relation 
in Fig.~\ref{FDR}(a), 
and confirm that it holds in our simulation.

When larger $E$ ($\neq 0$) is applied, 
$\langle I \rangle_{E,0}$ becomes nonlinear with $E$,
as shown in the inset of Fig.~\ref{FDR}(a).
In such NESSs, we find that the FDR is violated, i.e., 
for any $T$ that is independent of $\omega$, 
\begin{equation}
S_I(\omega;E) \neq 2 T {\rm Re}\tilde{\mu}(\omega;E)
\mbox{ for some $\omega$}.
\label{eq:FDR_neq}
\end{equation}
This is demonstrated in 
Fig.~\ref{FDR}(b), which
shows $S_I(\omega;E)$, $2 T_0 {\rm Re}\tilde{\mu}(\omega;E)$, 
and $2 T_{\rm e}(E) {\rm Re}\tilde{\mu}(\omega;E)$ 
in a nonlinear response regime.
Here, $T_{\rm e}(E) \equiv m_{\rm e} \langle (v^x_{\rm e} 
- \langle v^x_{\rm e} \rangle_{E,0} )^2 \rangle_{E,0}$
is a kinetic temperature of electrons  
($v^x_{\rm e}$ is the velocity of an electron in the $x$-direction).
When we employ $2 T_0 {\rm Re}\tilde{\mu}(\omega;E)$ 
as the right-hand side (RHS) of the FDR,
the violation of the FDR is observed 
in a wide frequency range.
When we use $2 T_{\rm e}(E) {\rm Re}\tilde{\mu}(\omega;E)$ as the RHS, 
the violation is observed at low frequencies 
($\omega \ll \omega_0$) \cite{note:Welch} 
while the RHS coincides with $S_I(\omega;E)$ at higher frequencies 
($\omega \gg \omega_0$),
where $\omega_0$ is the crossover frequency 
between the regimes of FDR violation and validation. 
These data also show that  
the FDR is violated 
for {\em any definitions of $T$} that is independent of $\omega$.

\begin{figure}[tb]
\begin{center}
\includegraphics[width=.84\linewidth]{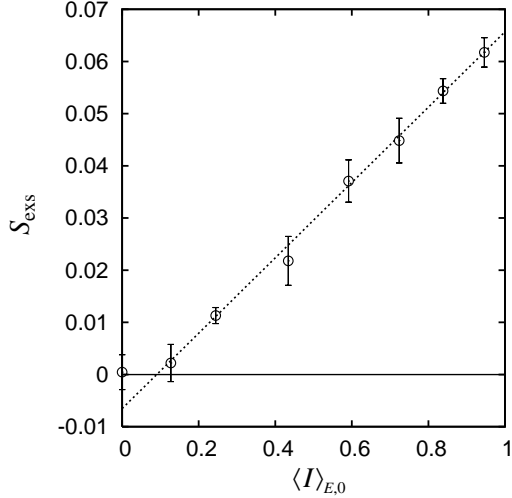}
\end{center}
\caption{
Excess fluctuation $S_{\rm exs}$ at a low frequency, 
plotted against $\langle I \rangle_{E,0}$.
The dotted line represents the asymptote, 
$W\bigl( |\langle I \rangle_{E,0}| - I_0 \bigr)$, 
fitted with the four data points at larger values of $\langle I \rangle_{E,0}$.
The parameters of the simulation and the meaning of the error bars
are the same as those in Fig.~\ref{FDR}.
}
\label{excess}
\end{figure}

Now we discuss the main finding of this paper.
Since we have seen that the FDR violation is manifested 
at lower frequencies, 
we look at the low-frequency region ($\omega \ll \omega_0$).
Among many possible definitions of 
`thermal fluctuation' of $I$ for $E \neq 0$, 
we employ 
\begin{equation}
S_{\rm th}(\omega;E) \equiv 2 T_0 {\rm Re}\tilde{\mu}(\omega;E),
\label{eq:thermal}
\end{equation}
which is the RHS of eq.~(\ref{eq:FDR_neq}) with $T=T_0$.
Using this, we decompose the total fluctuation $S_I$ 
into two parts: 
\begin{equation}
S_I (\omega; E) 
= S_{\rm th}(\omega;E) + S_{\rm exs}(\omega; E).
\end{equation}
Since the thus-defined $S_{\rm exs}$ quantifies the FDR violation, 
we call it excess fluctuation.
In Fig.~\ref{excess}, we plot $S_{\rm exs}$ for $\omega \simeq 0.002$ 
{\em as a function of $\langle I \rangle_{E,0}$}.
[We can translate a function of $E$
into a function of $\langle I \rangle_{E,0}$
because of the one-to-one correspondence 
between $E$ and $\langle I \rangle_{E,0}$.]
Since the FDR holds in equilibrium states, 
$S_{\rm exs} \simeq 0$  when $\langle I \rangle_{E,0}$ is small.
As $\langle I \rangle_{E,0}$ increases, 
$S_{\rm exs}$ exhibits a crossover behavior from 
near equilibrium to far from equilibrium as
\begin{equation}
S_{\rm exs} 
\simeq \left\{
  \begin{array}{ll}
    0  &  \bigl( |\langle I \rangle_{E,0}| \ll I_0 \bigr), \\
    W\bigl( |\langle I \rangle_{E,0}| - I_0 \bigr)
       &  \bigl( |\langle I \rangle_{E,0}| \gg I_0 \bigr), 
  \end{array}
\right.
\label{eq:asymptote}
\end{equation}
where $I_0$ is a certain crossover value of the current.
In the latter region 
($|\langle I \rangle_{E,0}| \gg I_0 $),
$S_{\rm exs}$ 
takes the form of shot noise,
where $W$ is the Fano factor
\cite{experiment1,BB00,SU92,BH91,KF,SS1991,SUS1993,BMS,book,HK1995,FS1998}.

\begin{figure}[tb]
\begin{center}
\includegraphics[width=.9\linewidth]{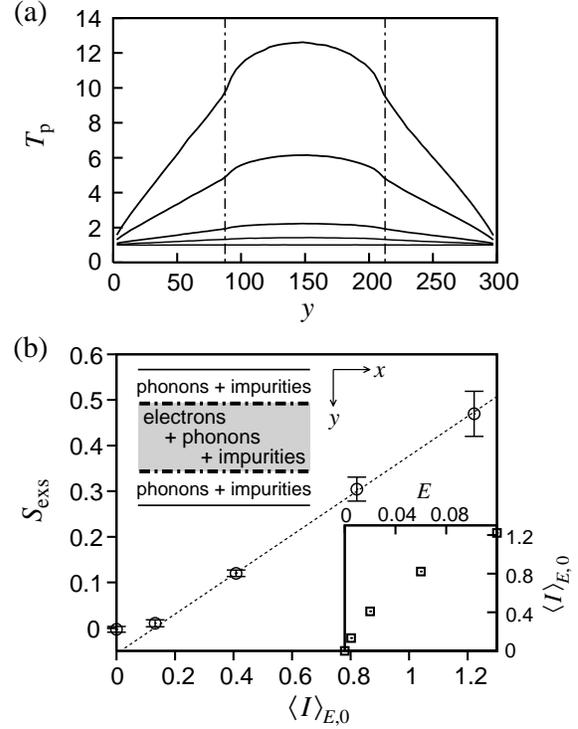}
\end{center}
\caption{
(a) Local phonon temperature $T_{\rm p}$ and
(b) excess fluctuation $S_{\rm exs}$ for $\omega \simeq 0.002$,
plotted against $\langle I \rangle_{E,0}$,
for a system where the thermal walls  (at $y=0$ and $300$)
for phonons are set 
away from the potential walls (at $y=87.5$ and $212.5$, 
dash-dotted lines) 
for electrons,
as shown in the top-left inset of (b).
In (a), 
the solid lines from bottom to top correspond to the data for
$E$=0,~0.01,~0.02,~0.06 and 0.12.
In (b), the dotted line represents the asymptote 
$W\bigl( |\langle I \rangle_{E,0}| - I_0 \bigr)$.
The meaning of the error bars 
is the same as that of the error bars in Fig.~\ref{FDR}.
Bottom-right inset: $\langle I \rangle_{E,0}$ versus $E$ for this system.
We take 
$m_{\rm p}=1$, $R_{\rm i}=0.5$, 
$L_x=375$, 
$n_{\rm e}=0.016$, 
$n_{\rm p}=19/1125$, and
$n_{\rm i}=2/375$.
}
\label{excess_i250i350}
\end{figure}

We have thus found that 
the dominant mechanism that breaks the FDR is 
the appearance of shot noise.
To confirm that this observation 
holds widely for the model considered here, 
we also study 
$S_{\rm exs}$ in the following cases: 
(i) another impurity density, $n_{\rm i}$ = 0.016, 
(ii) other linear dimensions $L_x$ (along $E$) of the system,
$L_x=375$, $300$, $187.5$, and $150$, and 
(iii) the values of the other parameters are changed significantly
(e.g., $n_{\rm e}=0.008$, $m_{\rm p}=10$, and $R_{\rm i}=2$).
(iv) The thermal walls for phonons are set 
away from the boundaries for electrons,
as shown in the top-left inset of Fig.~\ref{excess_i250i350}(b).

Figure \ref{excess_i250i350} shows the results in case (iv).
In this case, 
the local phonon temperature $T_{\rm p}$ 
around the boundaries for electrons 
is markedly different from 
$T_0$, as shown in Fig.~\ref{excess_i250i350}(a), where 
$T_{\rm p} \equiv m_{\rm p}\langle (v^x_{\rm p} - 
\langle v^x_{\rm p}\rangle_{E,0})^2\rangle_{E,0}$ 
($v^x_{\rm p}$ is the local phonon velocity in the $x$-direction).
Despite this fact, 
$S_{\rm exs}$ is well-fitted again by eq.~(\ref{eq:asymptote}),
as shown in Fig.~\ref{excess_i250i350}(b),
if we define thermal fluctuation again 
by eq.~(\ref{eq:thermal})
using $T_0$. 
Furthermore, we have found,
although the data are not shown here,
that eq.~(\ref{eq:asymptote}) also holds well in cases (i)-(iii)
(except when the densities are so high that a liquid-solid 
phase transition takes place).
Note in particular that the validity of eq.~(\ref{eq:asymptote}) in case (iii) 
suggests that it holds independently of details of the models,
because 
case (iii) naturally includes, for example, the case where 
the $R_j$ are specific functions of $n_{\rm e}$.

The above observations 
strongly indicate the robustness of eq.~(\ref{eq:asymptote}).
Note that this possible universality is visible only when 
thermal fluctuation in nonequilibrium states 
is appropriately defined as eq.~(\ref{eq:thermal}).
In fact, we have found (although the data are 
not shown here) 
that the possible universality is obscured if we use $T_{\rm e}(E)$ 
instead of $T_0$ in thermal fluctuation.

\begin{figure}[tb]
\begin{center}
\includegraphics[width=.85\linewidth]{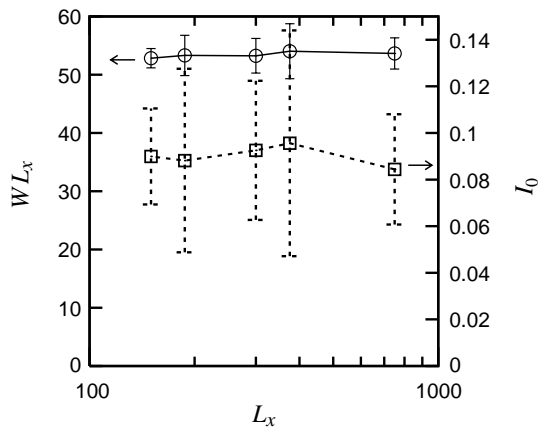}
\end{center}
\caption{
$WL_x$ (circles; left axis) and $I_0$ (squares; right axis) 
for several sizes of the system.
The particle densities are the same as those in Fig.~\ref{FDR}.
}
\label{WandI0}
\end{figure}

Using the results in case (ii), 
we also investigate the $L_x$ dependences of $W$ and $I_0$.
We evaluate $W$ and $I_0$ by fitting 
the numerical results of $S_{\rm exs}$ 
for large $\left| \langle I \rangle_{E,0} \right|$
to the asymptotic form of eq.~(\ref{eq:asymptote}).
In Fig.~\ref{WandI0}, we show $WL_x$ and $I_0$ versus $L_x$.
We see that $WL_x$ is almost independent of $L_x$, 
i.e., $W \sim 1/L_x$.
This agrees with the partial result 
for macroscopic conductors in ref.~\citen{SU92} 
(however, see note \cite{note:scaling}),
and coincides with the results for long mesoscopic 
conductors \cite{BB00,BH91}.
Furthermore, we observe that $I_0$ is almost independent of $L_x$,
although 
the error bars are somewhat large.

By combining the present results with 
the results on simple systems 
\cite{experiment1,BB00,SU92,BH91,KF,SS1991,Nagaev95,
JB95,Kogan,GMG2005,SUS1993,BMS,book,HK1995,FS1998}, 
we conjecture 
that the FDR is violated 
not in a random and system-dependent manner but 
in a universal manner by the appearance of shot noise,
for a wide range of systems from mesoscopic to macroscopic.
All details of individual systems are absorbed 
into $W$, $I_0$, and the differential response function 
${\rm Re}\tilde{\mu}$ (by which thermal fluctuation is defined).
The origin of current fluctuation 
in the present model is the chaotic behavior of 
interacting many particles in classical systems, 
while that in the simple systems 
\cite{experiment1,BB00,SU92,BH91,KF,SS1991,Nagaev95,
JB95,Kogan,GMG2005,SUS1993,BMS,book,HK1995,FS1998} 
is essentially the probabilistic nature of quantum 
or thermal-activation processes 
of noninteracting particles.
Despite such a marked difference,
$S_{\rm exs}$ takes an identical form in all systems. 
This observation may be used as a touchstone in 
nonequilibrium thermodynamics or statistical mechanics
beyond the linear response theory.

Note that the present results could never be obtained by 
a na\"ive perturbation expansion, in powers of the driving force $E$,
about an equilibrium state.
For example, the relation 
$S_{\rm exs} \propto \left| \langle I \rangle \right|$
suggests that such a power series would not converge for large $E$.
Using MD simulation, we have successfully investigated such 
a `non-perturbative regime.'
Our results may be confirmed experimentally, 
for example, in high-quality doped semiconductors, 
which may be prepared
by modulation doping, 
at room temperature.

In conclusion, we have presented
a study of excess fluctuations in a nonequilibrium system and 
found that the fluctuation-dissipation relation is violated 
in a manner that may be universal.
We hope that our work will stimulate further research that 
will test the correctness of this conjecture for 
wider classes of systems.

\section*{Acknowledgments}
The authors acknowledge N. Ito for helpful discussions.
T.Y. was supported by Research Fellowships 
of the Japan Society for the Promotion of Science for Young Scientists 
(No. 1811579).
This work was supported by KAKENHI No.~19540415 
and the Grant-in-Aid for the GCOE Program 
``Weaving Science Web beyond Particle-Matter Hierarchy''.


\begin{thebibliography}{99}
\bibitem{W_varies}
In some mesoscopic conductors, 
transmittance ${\cal T}$ 
varies as a function of applied voltage,
and so does $W$ ($=1-{\cal T}$).
Although $S_I$ then becomes a nonlinear function of 
$\left| \langle I \rangle \right|$,
it is also called shot noise because the physics involved is the same as 
that in the other cases where ${\cal T}$ is constant or small.
\bibitem{note:Welch}
In the two-sided Welch's test, the null hypothesis that 
$S_I(\omega \simeq 0;E=0.06) 
= 2 T_{\rm e} {\rm Re}\tilde{\mu}(\omega \simeq 0;E=0.06)$
is rejected at the 99.9\% confidence level.
\end{thebibliography}
\end{document}